\documentclass[10pt,conference]{IEEEtran}

\usepackage{cite}
\usepackage{amsmath,amssymb,amsfonts}
\usepackage{array}
\usepackage{graphicx}
\usepackage{algorithm}
\usepackage{algpseudocode}
\usepackage{textcomp}
\usepackage{xcolor}
\usepackage{tikz}
\usetikzlibrary{arrows.meta,backgrounds,calc,fit,positioning}
\usepackage[hyphens]{url}
\usepackage{fancyhdr}
\usepackage{hyperref}
\usepackage{cleveref} 
\usepackage{placeins}
\usepackage{enumitem}
\usepackage{pifont}
  \usepackage{array}
  \usepackage{booktabs}
  \usepackage{tabularx}
  \newcolumntype{Y}[1]{>{\hsize=#1\hsize\raggedright\arraybackslash}X}
\definecolor{llmgray}{gray}{0.35}

\newcommand{\hpcayear}{2027}
\newcommand{\hpcasubmissionnumber}{XXXX}

\title{\huge{HLSmith: An Expert-Guided Agentic Framework for C/C++-to-HLS Translation}}

\def\hpcacameraready{}

\newcommand\hpcaauthors{%
  Yuebo Luo\IEEEauthorrefmark{1},
  Ahmad Sedigh Baroughi\IEEEauthorrefmark{2},
  Philip Stachura\IEEEauthorrefmark{1}\IEEEauthorrefmark{2},
  Le Chen\IEEEauthorrefmark{3},
  Venkatram Vishwanath\IEEEauthorrefmark{3}, \\
  Zhenman Fang\IEEEauthorrefmark{1}\IEEEauthorrefmark{2},
  and Caiwen Ding\IEEEauthorrefmark{1}}
\newcommand\hpcaaffiliation{%
  \IEEEauthorrefmark{1}University of Minnesota Twin Cities \quad
  \IEEEauthorrefmark{2}Simon Fraser University \quad
  \IEEEauthorrefmark{3}Argonne National Laboratory}
\newcommand\hpcaemail{%
  \IEEEauthorrefmark{1}\texttt{luo00466@umn.edu, zhenman@umn.edu, dingc@umn.edu} \\
  \IEEEauthorrefmark{2}\texttt{asa582@sfu.ca, pstachur@sfu.ca} \quad
  \IEEEauthorrefmark{3}\texttt{lechen@anl.gov, venkat@anl.gov}}

\hypersetup{
  hidelinks,
  pdfauthor={Yuebo Luo, Ahmad Sedigh Baroughi, Philip Stachura, Le Chen,
    Venkatram Vishwanath, Zhenman Fang, and Caiwen Ding},
  pdftitle={HLSmith: An Expert-Guided Agentic Framework for C/C++-to-HLS Translation},
  pdfsubject={HPCA 2027 paper},
  pdfkeywords={high-level synthesis, FPGA, C-to-HLS translation,
  HLS optimization, compiler feedback, architectural expertise}
}

\newcommand{\coden}[1]{\texttt{#1}}
\newcommand{\system}{HLSmith}

\newenvironment{ps}{\begingroup\color{violet}\noindent Philip: }{\par\endgroup}

\definecolor{midgreen}{RGB}{0, 170, 0}
\newcommand{\greencheck}{\raisebox{0.2ex}{\textcolor{green!70!black}{\ding{51}}}}
\newcommand{\redcross}{\raisebox{0.2ex}{\textcolor{red!70!black}{\ding{55}}}}

\newcommand{\TRMMBaselineEstCycles}{22{,}598{,}401}
\newcommand{\TRMMCoalescingEstCycles}{469{,}321}
\newcommand{\TRMMDoubleBufferEstCycles}{181{,}621}
\newcommand{\TRMMPipelineEstCycles}{159{,}267}
\newcommand{\TRMMTilingEstCycles}{97{,}914}
\newcommand{\TRMMUnrollEstCycles}{26{,}233}

\newcommand{\TRMMFinalPeriod}{2.494}

\newcommand{\TRMMFinalRTLCycles}{50{,}040}
\newcommand{\TRMMOutputValues}{4{,}800}

\providecommand{\RouterDevelopmentKernelCount}{19}
\providecommand{\RouterExtraTreesExactCount}{17}
\providecommand{\RouterExtraTreesWithinFiveCount}{17}

\providecommand{\RouterExtraTreesTailRegret}{1.362}

\providecommand{\RouterFusionWithinFiveCount}{18}

\providecommand{\RouterFusionTailRegret}{1.087}
\providecommand{\RouterFusionPrunedPercent}{50.0}

\providecommand{\ModelMatrixAgenticValidCount}{60}
\providecommand{\GemmaDirectCSynthCount}{22}
\providecommand{\GemmaDirectKernelCount}{27}
\providecommand{\QwenBaseDirectCSynthCount}{21}
\providecommand{\QwenBaseDirectKernelCount}{27}

\providecommand{\TranslatorSFTABKernelCount}{4}
\providecommand{\TranslatorSFTABHLSGain}{1.70}
\providecommand{\TranslatorSFTABRTLGain}{1.60}
\providecommand{\TranslatorSFTABRTLCILow}{1.00}
\providecommand{\TranslatorSFTABRTLCIHigh}{2.56}

\providecommand{\TranslatorSFTABBaseTimingCount}{3}
\providecommand{\TranslatorSFTABTunedTimingCount}{4}

\providecommand{\NewSkillCommonKernelCount}{25}
\providecommand{\NewSkillCompletionCount}{180}
\providecommand{\NewSkillScheduledCount}{182}
\providecommand{\NewSkillAgenticCompletionCount}{154}
\providecommand{\NewSkillAgenticScheduledCount}{156}
\providecommand{\NewSkillAgenticFeasibleCount}{154}
\providecommand{\NewSkillBestGmean}{17.46}
\providecommand{\NewSkillNoSkillGmean}{6.07}
\providecommand{\NewSkillIncrementalGmean}{2.88}
\providecommand{\NewSkillEnabledWinCount}{22}

\providecommand{\NewSkillOneShotTimingCount}{18}

\providecommand{\GPTFiftyFiveCommonKernelCount}{23}

\providecommand{\GPTFiftyFiveAgenticAttemptCount}{153}
\providecommand{\GPTFiftyFiveAgenticExpectedCount}{156}
\providecommand{\GPTFiftyFiveAgenticCSynthPassCount}{152}
\providecommand{\GPTFiftyFiveAgenticFeasibleCount}{150}
\providecommand{\GPTFiftyFiveBestBaselineGain}{11.39}

\providecommand{\GPTFiftyFiveBestOverOffGain}{2.30}
\providecommand{\GPTFiftyFiveSkillsSelections}{17}

\providecommand{\SkillVThreeOpenCellCount}{540}

\providecommand{\SkillVThreeOpenKernelCount}{27}

\providecommand{\SkillVThreeOpenBaselineFallbackCount}{177}

\providecommand{\SkillVThreeOpenGemmaFlashReferenceReduction}{3.33}

\providecommand{\SkillVThreeOpenGemmaFlashOracleOverNoExpertise}{3.05}
\providecommand{\SkillVThreeOpenGemmaFlashExpertiseEnabledSelections}{16}

\providecommand{\SkillVThreeOpenGemmaMultistepReferenceReduction}{5.78}

\providecommand{\SkillVThreeOpenGemmaMultistepOracleOverNoExpertise}{4.71}
\providecommand{\SkillVThreeOpenGemmaMultistepExpertiseEnabledSelections}{20}

\providecommand{\SkillVThreeOpenQwenFlashReferenceReduction}{4.14}

\providecommand{\SkillVThreeOpenQwenFlashOracleOverNoExpertise}{18.77}
\providecommand{\SkillVThreeOpenQwenFlashExpertiseEnabledSelections}{23}

\providecommand{\SkillVThreeOpenQwenMultistepReferenceReduction}{2.98}

\providecommand{\SkillVThreeOpenQwenMultistepOracleOverNoExpertise}{11.39}
\providecommand{\SkillVThreeOpenQwenMultistepExpertiseEnabledSelections}{25}

\providecommand{\JulyThirtyDeepSeekPairedCount}{23}

\providecommand{\JulyThirtyDeepSeekGain}{3.22}

\providecommand{\JulyThirtyDeepSeekWins}{18}
\providecommand{\JulyThirtyDeepSeekLosses}{5}

\providecommand{\JulyThirtyGrokPrimaryCount}{23}
\providecommand{\JulyThirtyGrokPrimaryGain}{1.55}
\providecommand{\JulyThirtyGrokPrimaryWins}{18}
\providecommand{\JulyThirtyGrokPrimaryLosses}{5}

\providecommand{\JulyThirtyDeepSeekStrictCount}{22}
\providecommand{\JulyThirtyDeepSeekBaselineGain}{2.58}
\providecommand{\JulyThirtyDeepSeekStrictExpertiseGain}{2.55}
\providecommand{\JulyThirtyGrokStrictCount}{21}
\providecommand{\JulyThirtyGrokBaselineGain}{4.47}
\providecommand{\JulyThirtyGrokStrictExpertiseGain}{1.87}
\providecommand{\JulyThirtyGrokDeepSeekCommonCount}{18}
\providecommand{\JulyThirtyGrokOverDeepSeekGain}{1.66}

\providecommand{\JulyThirtyOpusBestBaselineGain}{14.51}

\providecommand{\JulyThirtyOpusBestOverOffGain}{2.63}
\providecommand{\JulyThirtyOpusGuidedSelections}{23}

\providecommand{\JulyThirtyOpusFlashSelectedOverOff}{1.03}
\providecommand{\JulyThirtyOpusFlashAllOverOff}{2.12}
\providecommand{\JulyThirtyOpusMultiSelectedOverOff}{1.30}
\providecommand{\JulyThirtyOpusMultiAllOverOff}{1.58}
\providecommand{\JulyThirtyHLSmithPass}{14}
\providecommand{\JulyThirtyChatHLSPass}{8}
\providecommand{\JulyThirtyChatHLSDenominator}{14}
\providecommand{\JulyThirtyChatHLSPaired}{8}
\providecommand{\JulyThirtyHLSmithOnlyPass}{6}
\providecommand{\JulyThirtyChatHLSFailureCount}{6}
\providecommand{\JulyThirtyChatHLSGain}{6.91}

\providecommand{\JulyThirtyChatHLSWithoutSyrTwoKGain}{4.24}

\providecommand{\GRPOMatchedKernelCount}{24}

\providecommand{\GRPOBaseRTLPassCount}{10}
\providecommand{\GRPOSFTRTLPassCount}{8}
\providecommand{\GRPORTLPassCount}{13}

\providecommand{\GRPOMedianDescriptiveRatio}{2.02}

\providecommand{\GRPOGesummvCycles}{4,987}
\providecommand{\GRPOGesummvReferenceCycles}{64,835}
\providecommand{\GRPOGesummvGain}{13.00}

\providecommand{\GRPOBrokenToCorrectCount}{7}
\providecommand{\GRPOBothCorrectFasterCount}{4}
\providecommand{\GRPOBaseCorrectRegressionCount}{3}

\author{
  \ifdefined\hpcacameraready
    \IEEEauthorblockN{\hpcaauthors{}}
      \IEEEauthorblockA{
        \hpcaaffiliation{} \\
        \hpcaemail{}
      }
  \else
    \IEEEauthorblockN{\normalsize{HPCA \hpcayear{} Submission
      \textbf{\#\hpcasubmissionnumber{}}} \\
      \IEEEauthorblockA{
        Confidential Draft \\
        Do NOT Distribute!!
      }
    }
  \fi 
}

\fancypagestyle{camerareadyfirstpage}{%
  \fancyhead{}
  \renewcommand{\headrulewidth}{0pt}
  \fancyhead[C]{
    \ifdefined\aeopen
    \parbox[][12mm][t]{13.5cm}{\hpcayear{} IEEE International Symposium on High-Performance Computer Architecture (HPCA)}    
    \else
      \ifdefined\aereviewed
      \parbox[][12mm][t]{13.5cm}{\hpcayear{} IEEE International Symposium on High-Performance Computer Architecture (HPCA)}
      \else
      \ifdefined\aereproduced
      \parbox[][12mm][t]{13.5cm}{\hpcayear{} IEEE International Symposium on High-Performance Computer Architecture (HPCA)}
      \else
      \parbox[][0mm][t]{13.5cm}{\hpcayear{} IEEE International Symposium on High-Performance Computer Architecture (HPCA)}
    \fi 
    \fi 
    \fi 
    \ifdefined\aeopen 
      \includegraphics[width=12mm,height=12mm]{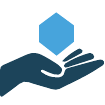}
    \fi 
    \ifdefined\aereviewed
      \includegraphics[width=12mm,height=12mm]{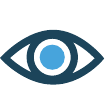}
    \fi 
    \ifdefined\aereproduced
      \includegraphics[width=12mm,height=12mm]{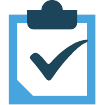}
    \fi
  }
  \fancyfoot[C]{}
}
\renewcommand{\headrulewidth}{0pt}

\begin{document}
\maketitle

\ifdefined\hpcacameraready 
  \thispagestyle{camerareadyfirstpage}
  \pagestyle{empty}
\else
  \thispagestyle{plain}
  \pagestyle{plain}
\fi

\newcommand{\hpcaheight}{0mm}
\ifdefined\eaopen
\renewcommand{\hpcaheight}{12mm}
\fi

\fancypagestyle{camerareadyfirstpage}{%
  \fancyhf{}
  \renewcommand{\headrulewidth}{0pt}
}


\begingroup
\begin{abstract}

Application-specific FPGA accelerators can provide substantial improvements in performance and energy efficiency across a range of application domains. However, developing FPGA designs is costly and often requires months of work due to the complexity and specialized expertise involved. Even with higher-level abstraction tools such as high-level synthesis (HLS), designers still require extensive hardware knowledge to develop high-performance accelerators.

Although large language models (LLMs) have demonstrated strong software-generation capabilities, even frontier models lack the hardware intuition and procedural knowledge needed to reliably translate baseline C/C++ programs into high-performance HLS designs. They struggle to identify effective architectures, follow the optimization process used by HLS experts, and apply hardware transformations consistently across diverse kernels.


We present \system{}, an expert-guided framework for translating C/C++ programs into optimized HLS accelerators. \system{} combines three components: an HLS optimization expertise library that encodes guarded transformation recipes, their applicability and prerequisite conditions, and unsafe cases to avoid; a staged, feedback-driven orchestration flow modeled on expert HLS development practice, which guides agents through synthesis, bottleneck analysis, and optimization; and a tool-grounded model-adaptation pipeline that converts optimization trajectories produced by commercial frontier models into training data for fine-tuning open-weight LLMs.

We evaluate \system{} on PolyBench and 
compare against ChatHLS, a leading prior agent-orchestration framework for HLS accelerator development. \system{} achieves a geometric mean speedup of 4.24x over ChatHLS while producing functionally correct designs, in both software and RTL simulation, for every benchmark compared with a 57\% valid-design rate of ChatHLS. And a speedup of up to 252x and 138x on PolyBench with commercial frontier models and open weight models, respectively.

\end{abstract}








\section{Introduction}
\label{sec:introduction}

Application-specific FPGA accelerators can provide substantial improvements in performance and energy efficiency by customizing the compute datapath, memory hierarchy, and parallelism to the requirements of an application~\cite{cong2011hls,coussy2009introduction}. However, these benefits come at the cost of long development cycles and substantial hardware-design expertise. A designer must translate the algorithm into an effective hardware architecture, making coupled decisions about parallelism, data movement, memory organization, and pipelining under resource and timing constraints. This makes FPGA accelerator development difficult even for experienced designers.

High-level synthesis (HLS) raises the design abstraction by compiling C/C++ programs into register-transfer level (RTL) hardware. HLS removes much of the burden of explicitly describing cycle-level control and datapaths, but it does not remove the need for hardware architecture design. Ordinary algorithmic C/C++ primarily specifies \emph{what} to compute; it rarely provides the source structure needed to express \emph{how} the computation should be implemented efficiently on an FPGA. 
\begin{figure*}[t]
  \centering
  \includegraphics[width=0.99\textwidth]{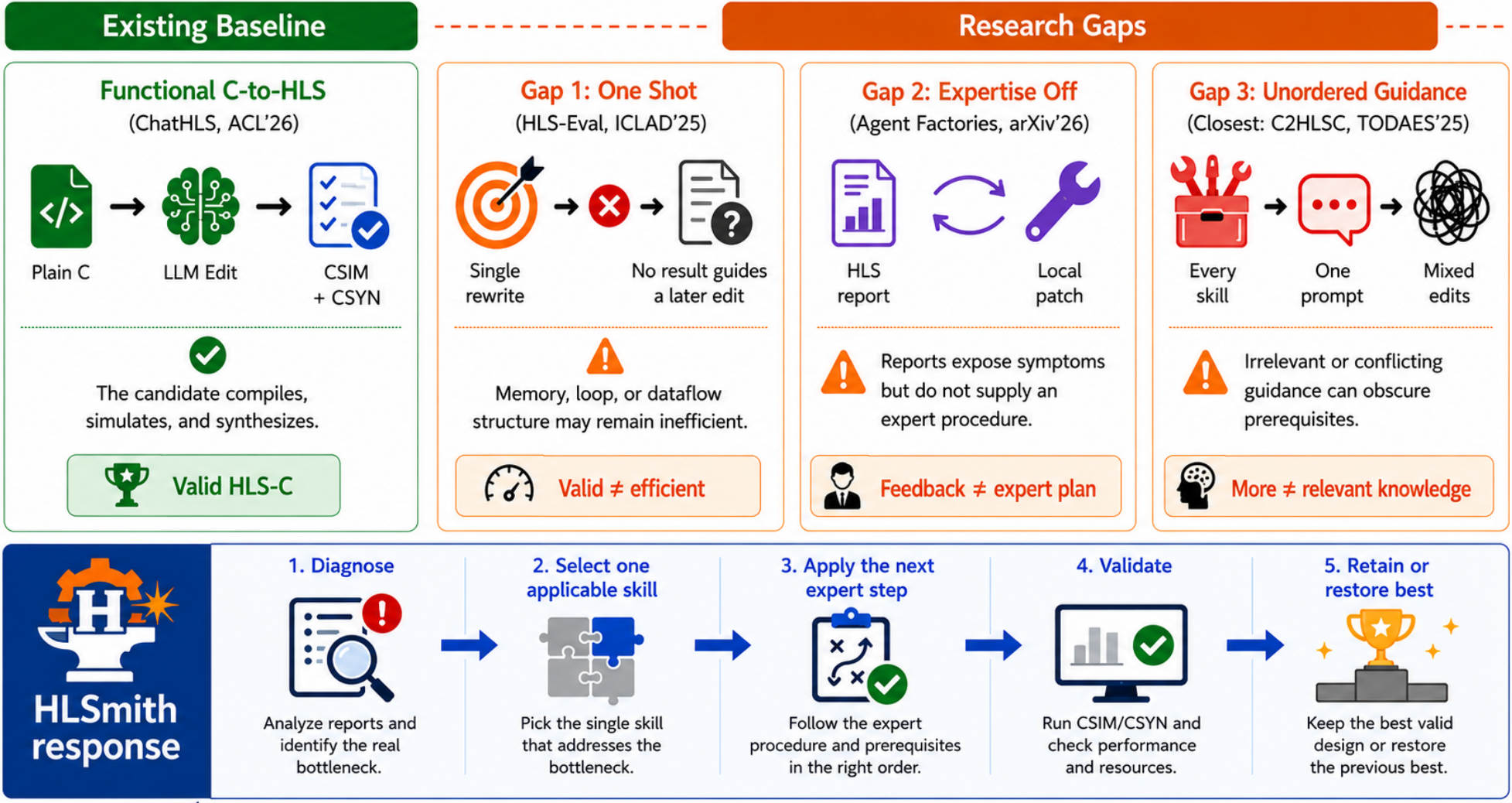}
  \caption{
  From functional HLS-C to architecture-aware optimization.  The top
  row isolates one-shot generation, feedback without explicit HLS expertise,
  and broad unordered guidance; these are control mechanisms rather than full
  characterizations of the cited systems.  The bottom row shows \system{}'s
  bottleneck-conditioned, prerequisite-aware, validated workflow.  Baseline
  exemplars are HLS-Eval, Agent Factories, C2HLSC, and
  LAAFD~\cite{abikaram2025hlseval,bhandwaldar2026agentfactories,
  collini2024c2hlsc,moraru2026laafd}; layout inspired by
  CUDAHercules~\cite{li2026cudahercules}.}
  \label{fig:baseline-gaps}
\end{figure*}

A high-performance HLS design commonly requires coordinated transformations such as creating tiled on-chip buffers, reorganizing loops for pipelining, partitioning memories to feed unrolled compute lanes, separating load, compute, and store stages, overlapping these stages with double buffering, and coalescing external-memory transfers~\cite{cong2018besteffort,sohrabizadeh2020autodse,pouget2025sisyphus}. This is not limited to simply applying a pragmas to an existing C/C++ implementation but often requires restructuring of the code to develop a properly architected and optimized design. Consequently, translating baseline C/C++ into optimized HLS requires the same fundamental hardware reasoning and iterative development process that limits conventional FPGA design.
Recent work has explored large language models (LLMs) for automating HLS code generation, compatibility repair, pragma selection, and optimization with synthesis feedback~\cite{abikaram2025hlseval,gai2025hlsmodels,collini2024c2hlsc,xiong2024hlspilot,xu2026hlsrewriter,Xu2024OptimizingModels,li2026chathls,mashnoor2025timelyhls,prakriya2025lift,moraru2026laafd,bhandwaldar2026agentfactories,zou2026agrefactor,zou2026hlsseek}. These systems demonstrate that LLMs can often produce code that compiles, passes C simulation, and synthesizes. However, a functionally valid HLS program is not necessarily a high-performance accelerator. It may repeatedly access external memory, serialize computation through a poorly selected loop, require more concurrent accesses than the available local-memory ports, or lack the architectural stages needed to overlap communication and computation. 

ChatHLS, the most versatile prior framework, specializes models for HLS-C generation, error repair, directive planning, and validation, but its optimization flow focuses on directive tuning and does not perform algorithm-level C/C++ refactoring~\cite{li2026chathls}. 
Other agentic systems, including LAAFD, Agent Factories, and AgRefactor, improve HLS optimization through iterative feedback, larger agent searches, or knowledge accumulated across tasks~\cite{moraru2026laafd,bhandwaldar2026agentfactories,zou2026agrefactor}. However, they do not explicitly encode or systematically evaluate the hardware-architecture expertise and multistep development process that experienced HLS designers use to progressively restructure baseline C/C++ into a high-performance accelerator.


We identify three limitations that prevent existing LLM-based approaches from consistently translating baseline C/C++ into expert-quality HLS accelerators, shown in Figure~\ref{fig:baseline-gaps}. 

\noindent (1) \textbf{General-purpose LLMs have weak hardware-architecture intuition}. They are effective at preserving software functionality but often fail to reason about the spatial architecture of the source code, including locality, banking, pipeline initiation interval, resource replication, and communication bandwidth. 

\noindent (2) \textbf{LLMs lack procedural knowledge of the HLS development process}. Experienced HLS designers do not treat C-to-HLS conversion as a single code-generation task. They follow a staged development process: first establishing a correct and synthesizable baseline, then diagnosing architectural bottlenecks, and finally refining the design through successive optimization stages that build on earlier transformations.

\noindent (3) \textbf{LLMs struggle to apply optimization strategies consistently across diverse kernels}. HLS transformations often depend on program structure and prerequisite changes: unrolling requires sufficient memory banking to feed replicated lanes, while double buffering requires separable load, compute, and store stages. A broad, unordered list does not tell the model which optimization is applicable or how to apply it safely.

\begin{table*}[t]
  \centering
  \caption{Methodological scope and public-source availability of representative
  LLM-based HLS systems.}
  \label{tab:system-scope}

  \scriptsize
  \setlength{\tabcolsep}{2.2pt}
  \renewcommand{\arraystretch}{1.08}

  \begin{tabular}{@{}
    >{\raggedright\arraybackslash}p{0.145\textwidth}
    *{4}{>{\centering\arraybackslash}p{0.082\textwidth}}
    |
    *{4}{>{\centering\arraybackslash}p{0.104\textwidth}}
    @{}}

    \hline
    &
    \multicolumn{4}{c|}{\textbf{Comparable end-to-end workflow scope}} &
    \multicolumn{4}{c}{\textbf{Optimization mechanism and availability}} \\

    \cline{2-5}\cline{6-9}

    \textbf{System} &
    \shortstack{\textbf{HLS-C}\\\textbf{generation}} &
    \shortstack{\textbf{Tool-grounded}\\\textbf{repair}} &
    \shortstack{\textbf{QoR-feedback}\\\textbf{optimization}} &
    \shortstack{\textbf{C/RTL}\\\textbf{verification}} &
    \shortstack{\textbf{Architecture}\\\textbf{restructuring}} &
    \shortstack{\textbf{HLS-specific}\\\textbf{model adaptation}} &
    \shortstack{\textbf{Guarded, ordered}\\\textbf{optimization recipes}} &
    \shortstack{\textbf{Open}\\\textbf{source}} \\

    \hline

    HLS-Eval (One-shot)~\cite{abikaram2025hlseval} &
      \greencheck & --- & --- & $\circ$ &
      $\circ$ & --- & --- & \greencheck \\

    C2HLSC~\cite{collini2024c2hlsc} &
      \greencheck & \greencheck & $\circ$ & $\circ$ &
      $\circ$ & --- & --- & \greencheck \\

    HLSRewriter~\cite{xu2026hlsrewriter} &
      \greencheck & \greencheck & $\circ$ & \greencheck &
      \greencheck & --- & $\circ$ & \redcross \\

    Agent Factories~\cite{bhandwaldar2026agentfactories} &
      $\circ$ & $\circ$ & \greencheck & $\circ$ &
      \greencheck & --- & --- & \redcross \\

    LAAFD~\cite{moraru2026laafd} &
      \greencheck & \greencheck & \greencheck & \greencheck &
      \greencheck & --- & --- & \redcross \\

    AgRefactor~\cite{zou2026agrefactor} &
      \greencheck & \greencheck & \greencheck & $\circ$ &
      \greencheck & --- & $\circ$ & \greencheck \\

    \textbf{ChatHLS}~\cite{li2026chathls} &
      \greencheck & \greencheck & \greencheck & \greencheck &
      --- & \greencheck & --- & \greencheck \\

    \hline

    \textbf{\system{} (this work)} &
      \greencheck & \greencheck & \greencheck & \greencheck &
      \greencheck & \greencheck & \greencheck & Pending \\

    \hline
  \end{tabular}

  \vspace{1mm}
  \begin{minipage}{0.95\textwidth}
    \footnotesize
    $\circ$: partial;
    ---: absent.
  \end{minipage}
\end{table*}


In this paper, we present \system{}, an expert-guided framework for translating baseline C/C++ programs into optimized HLS accelerators. \system{} encodes HLS optimization expertise through a structured library of guarded transformation recipes and avoidance rules. Each library entry associates a bottleneck with structural preconditions, ordered code changes, and safety guards. These recipes and rules guide a staged, feedback-driven orchestration flow that first establishes a correct and synthesizable HLS baseline, then uses HLS tool reports to diagnose performance bottlenecks and apply successive optimizations when their preconditions are met. The same tool-grounded workflow records optimization trajectories produced by commercial frontier models and converts them into training data for fine-tuning open-weight LLMs. Together, these components capture both reusable HLS optimization expertise and the staged development process required to produce high-performance accelerators.


We evaluate \system{} on HLSFactory-28, a fixed set of 28 PolyBench kernels.
Compared with ChatHLS, \system{} achieves an average speedup of 4.2$\times$ while producing functionally correct designs in both software and RTL simulation, compared with a 57\% valid-design rate for ChatHLS. Fine-tuning further improves performance by up to 3.61$\times$ for a single kernel and $70\%$ overall improvement with GRPO applied to models as small as 7B, and with commercial models achieves up to 252$ \times$ speedup over the reference HLS kernel. 



In summary, this paper makes the following contributions:
\begin{enumerate}[ label=\arabic*),
    leftmargin=1.6em,
    labelsep=0.45em,
    itemsep=0.25em,
    topsep=0.25em
]
    \item A structured HLS optimization-expertise library comprising guarded transformation recipes and avoidance rules that encode bottleneck and applicability conditions, relevant structural requirements, ordered transformation steps, and safety guards.
    \item A staged, feedback-driven orchestration framework modeled on expert HLS development practice, which separates baseline synthesis, tool-guided bottleneck analysis, and successive optimization into distinct phases, greatly improving LLM actors' performance from zero-shot attempts.
    \item A tool-grounded model-adaptation methodology that converts commercial frontier-model optimization trajectories, evaluated using correctness and HLS-tool feedback, into training data for fine-tuning and reinforcement learning on open-weight LLMs to perform HLS optimization across diverse kernels.
\end{enumerate}

\begin{figure*}[t]
    \centering
    \includegraphics[
        width=\textwidth,
    ]{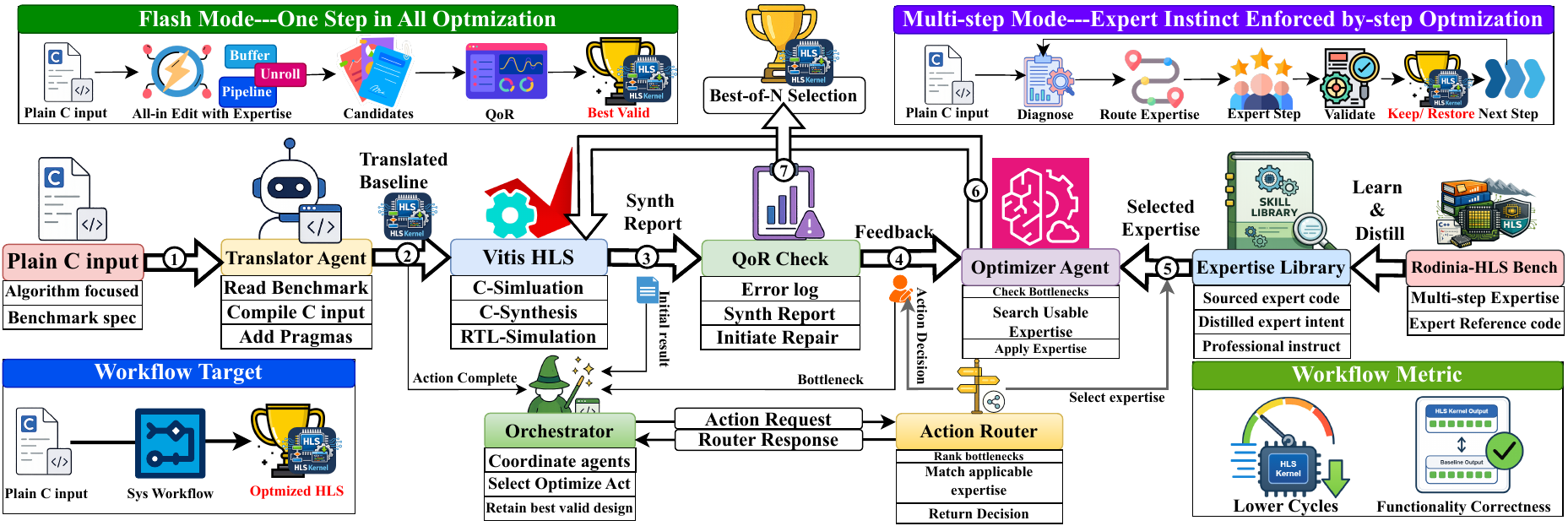}
    \caption{End-to-end \system{} workflow.  The numbered
     path translates plain C, evaluates the HLS-C design, diagnoses its
     bottleneck, routes the guarded recipe in Eq.~\eqref{eq:expert-skill}, and
     validates the resulting edit.  The upper paths contrast independent
     single-step candidates with repeated multi-step feedback; the
     Orchestrator coordinates both and preserves the best eligible design.}
    \label{fig:system}
    \vspace{-10pt}
\end{figure*}

\section{Related Work}
\label{sec:related}
Traditional compiler- and DSE-based approaches automate HLS optimization through synthesis-guided search, multi-level transformations, and joint optimization of code and directives~\cite{sohrabizadeh2020autodse,ye2021scalehls,pouget2025sisyphus}; however, they do not investigate reusable optimization expertise as an explicit inference-time component of an LLM workflow.

\textbf{LLM-based HLS generation, evaluation, and optimization.}
Recent work has established HLS as a distinct target for language models. HLS-Eval~\cite{abikaram2025hlseval} provides a benchmark and evaluation framework for two core tasks: generating HLS code from natural-language descriptions and editing HLS code for optimization, while Gai \emph{et al.}~\cite{gai2025hlsmodels} study benchmark construction, fine-tuned HLS generation, and prompting effects such as chain-of-thought and feedback loops. HLSPilot~\cite{xiong2024hlspilot} reduces the gap between sequential C/C++ and HLS by supplying pattern-specific C-to-HLS strategies through in-context learning, then relying on profiling and downstream DSE for pragma tuning. RALAD~\cite{Xu2024OptimizingModels} provides an existing, non-HLS-specialized LLM with retrieved HLS code examples and explanations to guide pragma insertion. LIFT~\cite{prakriya2025lift} trains an LLM to infer predefined pragma configurations, whereas HLS-Seek~\cite{zou2026hlsseek} trains an HLS code generator with QoR rewards and selective synthesis feedback.

\textbf{LLM-based refactoring and repair for HLS.}
A second line of work focuses on transforming software-oriented C/C++ into synthesizable or better-optimized HLS-C. C2HLSC~\cite{collini2024c2hlsc} iteratively refactors generic C into HLS-compatible C using hierarchical preprocessing and feedback from compilation and HLS, then performs a separate pragma-identification stage. HLSRewriter~\cite{xu2026hlsrewriter} starts from regular C/C++ and combines step-wise compatibility analysis, a RAG-based repair-template library extracted from HLS manuals, pipeline-aware loop decomposition, bit-width adjustment, and subsequent pragma-based PPA optimization. TimelyHLS~\cite{mashnoor2025timelyhls} uses a structured, architecture-specific knowledge base and iteratively refines generated HLS code using HLS and RTL synthesis logs, timing reports, and functional-correctness feedback. These systems show that LLMs can assist with compatibility repair, code restructuring, and architecture-aware refinement, but they do not isolate the effect of explicitly represented, reusable HLS optimization expertise combined with guarded and ordered application of transformations.

\textbf{Agentic and feedback-driven HLS workflows.}
ChatHLS~\cite{li2026chathls} is the most directly comparable reproducible baseline for our study. It provides a multi-agent workflow spanning HLS-C generation, tool-grounded debugging, verification, and iterative QoR-aware directive tuning. Its optimization component, HLSTuner, models the directive-to-hardware-to-QoR relationship and selects, configures, and inserts directives such as \textsc{Pipeline}, \textsc{Unroll}, and \textsc{Array\_Partition}, without performing source-level architectural restructuring.

LAAFD~\cite{moraru2026laafd} uses iterative synthesis feedback within an agentic workflow for latency-oriented HLS kernel generation and optimization. Agent Factories~\cite{bhandwaldar2026agentfactories} explores inference-time agent scaling through sub-kernel decomposition, ILP-based composition, and design-wide refinement. AgRefactor~\cite{zou2026agrefactor} combines cross-task memory with automated refactoring tools to ensure HLS compatibility and optimize performance. In particular, it integrates HeteroRefactor, a profile-guided, source-to-source refactoring and optimization tool for FPGA HLS, rather than a complete compiler or HLS backend. Although these systems employ tool feedback, multiple agents, source transformations, or reusable memory, they do not explicitly isolate the effect of guarded, ordered HLS optimization recipes across successive transformations.

\textbf{Comparison scope and artifact availability.}
Table~\ref{tab:system-scope} compares methodological scope and public-source availability rather than reported QoR, since the systems use different benchmarks, hardware targets, optimization objectives, and execution overhead. In the table, tool-grounded repair means that an HLS-tool failure conditions a subsequent code edit, while QoR-feedback optimization means that measured latency or resource results condition a subsequent optimization. Architecture restructuring denotes source-level changes to loop organization, memory access, buffering, or dataflow structure beyond directive insertion. HLS-specific model adaptation denotes fine-tuning or other changes to model parameters using HLS-specific data. Guarded, ordered optimization steps are reusable representations that encode applicability conditions, guards, and an ordered sequence of transformation steps.



\textbf{Positioning of HLSmith.}
HLSmith studies a mechanism not explored by prior LLM-based HLS systems. Rather than asking only whether retrieval helps, whether general-purpose agents can scale, or whether iterative feedback improves optimization, we study whether HLS optimization expertise can be represented as reusable, guarded procedures and applied through a staged controller that accounts for transformation order and applicability. HLSmith represents this expertise as a library of guarded transformation recipes and avoidance rules. Each entry encodes a bottleneck context, applicability conditions, relevant structural requirements, safety guards, and ordered transformation steps. This representation separates three mechanisms that are often conflated in prior work: selecting relevant HLS knowledge, applying a source-level architectural transformation, and determining which subsequent optimization is applicable after the current structural changes have been established. We further compare a single-turn Flash flow with a structured multi-step flow under matched controls, evaluate candidates using the HLS tool, and retain the best valid candidate design.

\section{C-to-HLS Task and Objective}
\label{sec:task}


\subsection{Input, Output, and Evaluation Setup}

For each kernel, \system{} receives an algorithmic C implementation without HLS pragmas, a corresponding reference HLS implementation used as the gold reference, the public header and top-level function interface, test inputs, independently computed CPU reference outputs, and the target hardware configuration and resource constraints. \system{} then produces an optimized C/C++ implementation for Vitis HLS. The generated implementation may restructure loops and arrays, introduce local buffers, add explicit processing stages, reorganize memory accesses to enable burst transfers, and insert HLS pragmas for interfaces, pipelining, loop unrolling, array partitioning, dataflow, and AXI memory configuration.

\subsection{Candidate Eligibility, Selection, and Validation}
For a method $m$, let $\mathcal{C}_m$ be the candidates generated within its resource budget.  The eligible set $\mathcal{E}_m\subseteq\mathcal{C}_m$ contains only
candidates that pass C simulation with verified functional correctness,
complete C synthesis, fit every target-device resource limit, and meet the
target clock period.  
The orchestrator (\Cref{sec:methodology}.A) selects based on~\Cref{eq:selection} as follows:
\begin{equation}
  c_m^\star=\arg\min_{c\in\mathcal{E}_m}\widehat L(c),
  \label{eq:selection}
\end{equation}
where $\widehat L(c)$ is the worst-case HLS-synthesis latency.  If
$\mathcal{E}_m$ is empty, the method fails on the task.  Only selected candidates $c_m^\star$
undergoes RTL co-simulation.

\section{Methodology}
\label{sec:methodology}

\begin{figure*}[t]
  \centering
  \includegraphics[
    width=\textwidth,
    trim={50pt 258pt 70pt 42pt},
    clip
  ]{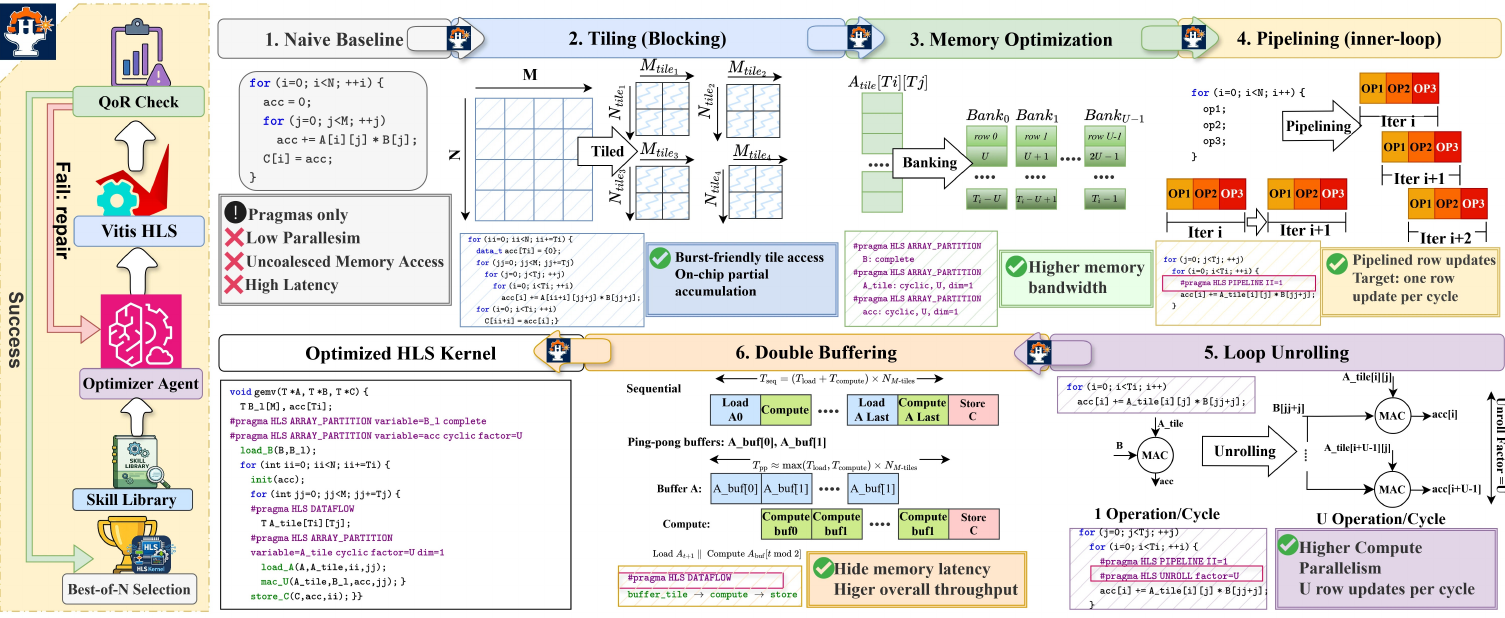}
  \caption{{Prerequisite-aware HLS optimization.  Tiling
  creates local reuse, banking supplies ports for parallel lanes, and stage
  separation enables double-buffered overlap.  Each step instantiates the
  contract in Eq.~\eqref{eq:expert-skill}.}}
  \label{fig:expert-prior}
\end{figure*}

\subsection{Agent Workflow and Compiler Feedback}
\label{sec:system}
\system{} assigns distinct roles to translation,
tool-grounded evaluation, bottleneck diagnosis, and optimization.  The agents
are execution mechanisms.

\textbf{Translator.}  At step \textcircled{1} in
Fig.~\ref{fig:system}, the Translator converts algorithmic C into an initial HLS-C implementation.

\textbf{Toolchain evaluator.}  Vitis C simulation and
synthesis establish functionality, synthesizability, latency, timing,
and used hardware resources.  

\textbf{QoR Check.}  The Feedback role converts tool output
into a concise diagnosis: compile or correctness faults, performance
bottlenecks involving latency, initiation interval, timing, resource pressure,
dependencies, scheduling, and memory ports.  
This diagnosis identifies the
triggering bottleneck $b$ used by the expertise in
Eq.~\eqref{eq:expert-skill}.
\textbf{Optimizer.}  The Optimizer receives the diagnosis profile and selects useful expertise from the expertise library.  Execute the candidate HLS kernel rewrite following expertise guidance.  A
multi-step run repeats this loop up to a default cap of five attempts.
Final selection
uses Eq.~\eqref{eq:selection}, and only the selected winner proceeds to the
reported RTL check.


\textbf{Orchestrator.}  The Orchestrator sequences the aforementioned roles, maintains the current and best eligible designs, and routes the best configuration combination set by flash mode/multi-step mode execution and expertise policies.  It asks the Action Router for expertise that matches the current bottleneck diagnosis, evaluates the resulting candidate, and either advances the search or restores the previous best state.


\system{} supports two search depths.  In
\emph{flash} mode, independent candidates receive the same initial
report, and each attempts one expertise-guided code generation.  In \emph{multi-step}
mode, every accepted candidate is evaluated again, and the new evidence
conditions the next recipe.  This separates one-turn guidance from
prerequisite-aware sequencing and repeated feedback.


Fig.~\ref{fig:system} shows an overview of the system framework.  The path in the middle illustrates how the system generally translates (\textcircled{1}), profiles the initial version (\textcircled{2},\textcircled{3} and \textcircled{4}), then optimizes HLS candidates with sourced expertise from the library (\textcircled{5} and \textcircled{6}) until the final candidate wins (\textcircled{7}); the lower path shows the Orchestrator and Action Router controlling the loop.  
The top left and right panels contrast
all-in-one flash mode with expertise-enforced multi-step optimization mode.  
\subsection{From Rodinia-HLS Patterns to Optimization Expertise}
\label{sec:search}

\begin{figure*}[t]
  \centering
  \includegraphics[width=\textwidth]{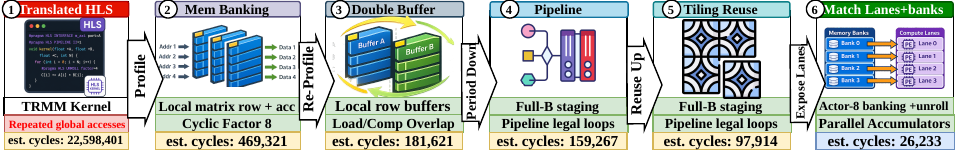}
  \caption{HLS-Factgory's TRMM optimization process with \system{} (base model Claude Sonnet 4.6).  Each intermediate design synthesizes and lowers Vitis's cycle
  estimate, with the CO-SIM measurement of the final design.  This illustrates the multi-step mechanism that builds up an overall improvement.
  }
  \label{fig:case-study}
\end{figure*}

Rodinia-HLS~\cite{cong2018fpga_gpu} supplies abundant and staged human HLS
optimization behaviors.
It provides examples of step-by-step architectural changes including tiling intermediate variables, pipelining legal loops, matching compute lanes to memory
banks, overlapping separable stages, and coalescing data transfer.  A specific HLS kernel design can integrate several optimization techniques, so \system{} learns and distills the fundamental optimization principles rather than the individual changes of specific HLS kernels in the benchmark.

\subsubsection{Implementation of Per-step Expertise}

Each set of optimization techniques applied is represented as:
\begin{equation}
  s=\langle b,p,a,q,g,e\rangle,
  \label{eq:expert-skill}
\end{equation}
where $b$ is a bottleneck found in the current HLS kernel diagnosis profile, $p$ the HLS code structural
prerequisites, $a$ the architectural action such as rewrite and edit, $q$ an ordered action checklist,
$g$ unsafe conditions that the actions should avoid, and $e$ the expected results from the actions.
Figure~\ref{fig:expert-prior} maps these fields to a prerequisite-aware sequence:
the diagnosis supplies $b$; required structures supply $p$; the depicted
transformations and arrows encode $a$ and $q$; legality and resource constraints
form $g$; and expected reuse, port, and overlap improvements form $e$.
Equation~\eqref{eq:expert-skill} is the interface between
expert knowledge and the agentic workflow.  At turn $t$, the QoR Check
produces bottleneck $b_t$.  The Action Router admits a recipe of actions to optimize only when its
$b$ matches $b_t$, its prerequisites $p$ hold in the current design, and no guard $g$ is violated.  The Optimizer receives the optimization action $a$ and checklist $q$;
the Vitis report generated from the optimized candidate is compared with $e$.
The previous best design is restored on failure; otherwise,
Eq.~\eqref{eq:selection} chooses the final eligible candidate.

In the figure, tiling creates reusable local data, banking supplies ports before
unrolling creates parallel lanes, and stage separation enables double-buffered
overlap.  The order is conditional rather than fixed: each QoR report may skip
or reorder actions whose prerequisites hold and whose guards are not violated.


\begin{figure}[t]
  \centering
  \includegraphics[width=\columnwidth]{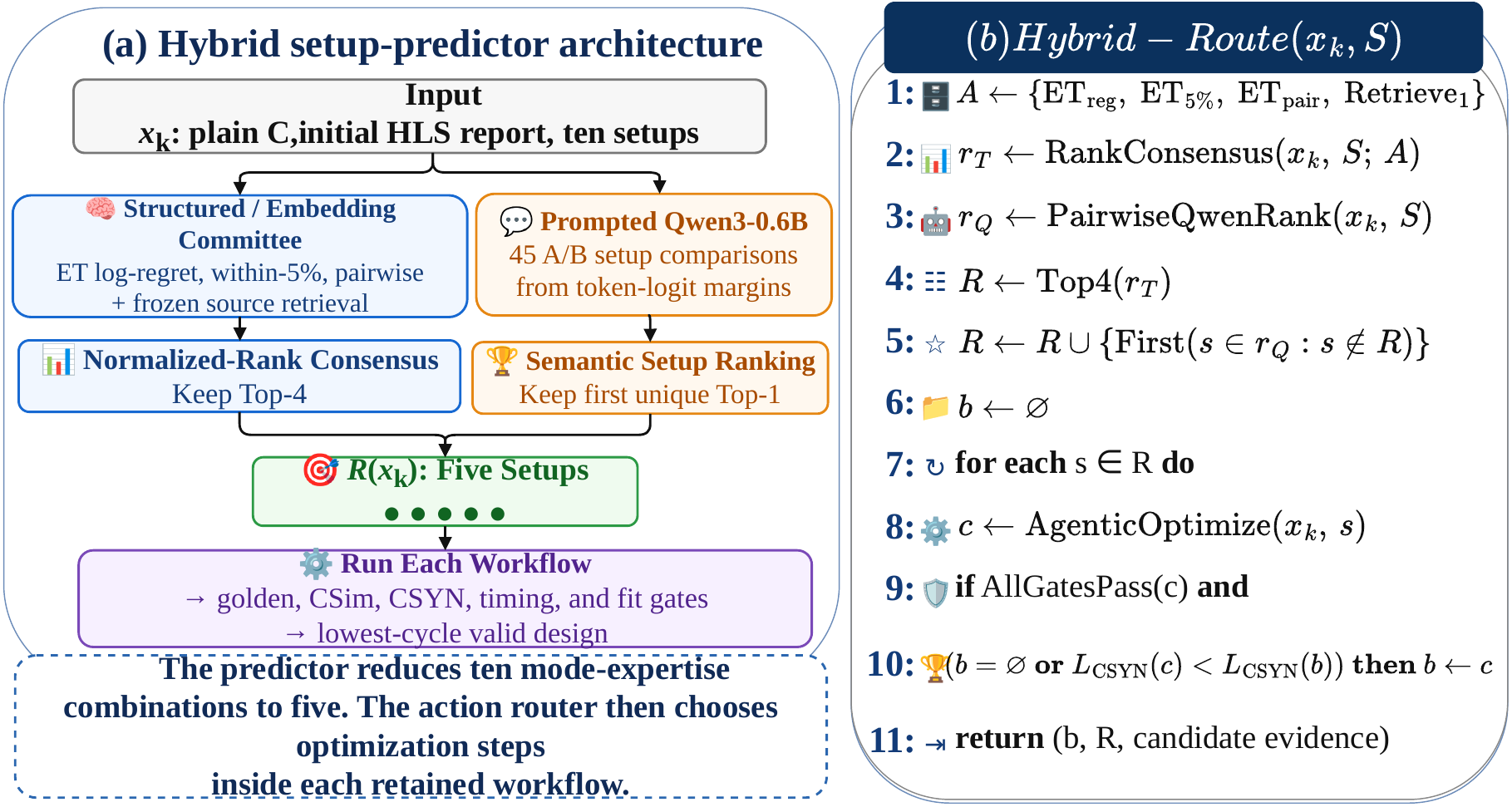}
  \caption{Hybrid setup routing.  A structured rank committee and an
  independent semantic comparator rank workflow configurations before
  candidate generation; fusion retains complementary choices for execution.}
  \label{fig:hybrid-router}
\end{figure}

\subsection{Expertise-Guided Optimization and Best-Design Recovery}

\textcolor{black}{The Action Router maps each diagnosed bottleneck to one of five source-level optimization families: local tiling, pipelining, unrolling with memory banking, double-buffered stage overlap, and memory-access coalescing. At each optimization step, the router ranks applicable library entries and provides the selected entry's applicability pattern, strategy, ordered transformation steps, and safety guards to the Optimizer Agent. In multi-step execution, each new synthesis report triggers another diagnosis and routing step, allowing subsequent transformations to reflect the current implementation and its remaining bottlenecks. Under the selected-expertise, the Optimizer receives only the routed entries, whereas the all-expertise provides all library entries and therefore evaluates targeted selection against broad exposure to expertise.}

Each generated candidate is first compared with the independent functional reference for functional correctness checks. Candidates that pass these checks undergo synthesis to determine latency, timing, and resource feasibility. A failed, unchanged, or lower-quality candidate does not replace the best eligible design that the Orchestrator retains. When optimization terminates, the internal scores used to manage intermediate steps are discarded, and Eq.~\eqref{eq:selection} selects the eligible candidate with the lowest HLS-estimated latency. Only this selected design undergoes RTL co-simulation; a run is considered valid, and RTL cycles are reported, only if co-simulation completes successfully and its outputs match the same independent reference.

\begin{algorithm}[t]
  \caption{Vitis-grounded GRPO adaptation}
  \label{alg:vitis-grpo}
  \footnotesize
  \begin{algorithmic}[1]
    \Require Tasks $\mathcal{D}=\{(x_k,L_k^0)\}$, resource limits $\mathbf B$,
      policy $\pi_\theta$, group size $G$, stabilizer $\epsilon_A$
    \For{each policy update}
      \For{each task context $x_k$ in the batch}
        \State Sample $a_{k,1},\ldots,a_{k,G}\sim
          \pi_{\theta_{\mathrm{old}}}(\cdot\mid x_k)$
        \For{$i=1,\ldots,G$ \textbf{in parallel}}
          \State $(z_{k,i},Q_{k,i})\gets
            \Call{VitisCSimCSynth}{a_{k,i},k}$
          \State $r_{k,i}\gets
            \Call{HLSReward}{z_{k,i},Q_{k,i},L_k^0,\mathbf B}$
        \EndFor
        \State $\mathbf A_k\gets
          (\mathbf r_k-\operatorname{mean}(\mathbf r_k))/
          (\operatorname{std}(\mathbf r_k)+\epsilon_A)$
      \EndFor
      \State Apply a completion-token-masked clipped GRPO update
    \EndFor
    \State \Return adapted Translator
  \end{algorithmic}
\end{algorithm}

\subsection{Example: Multi-Step Optimization of HLSFactory TRMM}
\label{sec:case-study}

\begin{table*}[t]
  \centering
  \caption{Router and post-training configurations.}
  \label{tab:router-training-settings}
  \scriptsize
  \setlength{\tabcolsep}{5pt}
  \renewcommand{\arraystretch}{1.12}
  \begin{tabularx}{\textwidth}{@{} l Y{0.95} Y{1.05} @{}}
    \toprule
    \textbf{Component} & \textbf{Model / data} & \textbf{Key settings} \\
    \midrule
    Early setup router &
    Qwen3-0.6B; 19/4/4/1 (train/validate/test/excluded) split; 874/191/186
    (train/validate/test) historical and 190/40/40 complete rows &
    ExtraTrees: latency ratio, within-5\%, and pairwise order; embedding
    retrieval; 45 pairwise prompts; structured top four $+$ unique semantic
    top one (five of ten setups) \\
    \addlinespace
    Supervised fine-tuning &
    Sonnet~4.6 $\rightarrow$ Qwen3.6-27B; 165 actions; 22 training/five
    held-out kernels; Translator, Synthesis, and Orchestrator &
    4-bit NF4 QLoRA~\cite{dettmers2023qlora}; rank 16; one epoch \\
    \addlinespace
    Group-relative policy optimization &
    Qwen2.5-Coder-7B; 14 Rodinia/ML4Accel training tasks; PolyBench evaluation &
    Group size 8; LoRA rank 32; clip 0.2; zero KL coefficient; temperature 1.0;
    1,536-token cap; 100 updates; learning rate $10^{-5}$ \\
    \bottomrule
  \end{tabularx}
\end{table*}

\begin{table}[t]
  \centering
  \caption{Router search space: two workflows $\times$ five expertise policies.}
  \label{tab:router-setup-space}
  \scriptsize
  \setlength{\tabcolsep}{3pt}
  \renewcommand{\arraystretch}{1.12}
  \begin{tabularx}{\columnwidth}{@{} l l Y{1} c @{}}
    \toprule
    \textbf{Expertise} & \textbf{Policy} & \textbf{Library view} &
    \begin{tabular}[b]{@{}c@{}}\textbf{Candidates}\\\textbf{flash / multi}\end{tabular} \\
    \midrule
    None & Disabled & Empty & 1 / 1 per step \\
    \midrule
    Selected & Matched & Up to three deterministic matches & 1 / 1 per step \\
    \cmidrule(l){2-4}
    & Best-fit & Zero to three evidence-ranked entries & 1 / 1 per step \\
    \cmidrule(l){2-4}
    & Exhaustive & One candidate per eligible entry (up to five) &
      $\leq5$ / $\leq5$ per step \\
    \midrule
    All & Full & 42 positive entries, applicability-gated & 1 / 1 per step \\
    \bottomrule
  \end{tabularx}
\end{table}

Using expertise from Rodinia-HLS to the HLS-Factory benchmark's TRMMkernel, the plain C code updates each row of a dense matrix from later rows and one strided column of a triangular matrix.  
Its direct loop nest repeatedly moves the same data and exposes neither sufficient reuse nor independent memory-fed arithmetic lanes.

Fig.~\ref{fig:case-study} shows how the \system{} translates and optimizes the TRMM kernel with base model Claude Sonnet 4.6.  The baseline kernel starts from a \TRMMBaselineEstCycles-cycle HLS estimate; after the \system{} applies local staging and cyclic banking, it reduces the estimate to \TRMMCoalescingEstCycles{} cycles.  On top of that, a double-buffered
candidate that stems from the previous step reaches \TRMMDoubleBufferEstCycles{} cycles but misses the 3.33-ns clock period target, so the next report changes the direction of optimization.  Full-matrix local
staging and a pipeline rewrite is then applied to reach \TRMMPipelineEstCycles{} cycles; then, the tiling is applied to the last best candidate
and reaches \TRMMTilingEstCycles{} cycles.  Finally, factor-eight banking and
unrolling match memory ports to arithmetic lanes are implemented on the last step's work and reach
\TRMMUnrollEstCycles{} cycles at \TRMMFinalPeriod{} ns. All five intermediate designs synthesize and improve the
preceding cycle estimate.  The selected design reports
\TRMMFinalRTLCycles{} executed co-simulation cycles, and all
\TRMMOutputValues{} output values match the independent CPU result. 



For prompt-time expertise comparisons, we hold the model weights fixed and vary only the expertise provided to the model. For model-adaptation comparisons, we hold the workflow and expertise policy fixed and vary only the model weights. Kernels used for evaluation are excluded from the model-adaptation training data.


\subsection{Hybrid Early-Prediction Setup Routing}

\label{sec:setup-router}
We define a \emph{workflow setup} as a complete HLSmith configuration.
Throughout the \system{}, the inner Action Router in Fig.~\ref{fig:system} selects an optimization action list \emph{within} a single workflow run.  A second, outer \emph{early-prediction setup router} decides \emph{which workflow configurations are worth proceeding}.
The setup router observes only the plain C source and its initial HLS report,
then ranks workflow-depth and expertise-policy combinations before the design is generated.  Section~\ref{sec:experimental-setup} specifies
the evaluated setup space and execution overhead in terms of time and tokens.

Fig.~\ref{fig:hybrid-router} shows the two complementary ranking branches.  The structured branch combines latency prediction,
near-optimal classification, pairwise ordering, and preference transfer from similar kernels.  Rank consensus merges these features.  
The semantic branch, supported by Qwen-0.6B, compares configuration descriptions in the context of the source and initial HLS report, then contributes a non-duplicate
recommendation; this prioritizes agreement among measured predictors while reserving an independent semantic judgment.
The concrete predictors, shortlist size, and data split can be found in
Section~\ref{sec:experimental-setup}.

Let $\mathcal{S}$ denote the available setups, $R(x_k)\subset\mathcal{S}$ the
subset retained for kernel context $x_k$, and $L_k(s)=\infty$ when setup $s$
does not produce a valid, timing- and resource-feasible design.  We quantify
the cost of pruning by the latency ratio as follows:
\begin{equation}
  \Gamma_k(R)=
  \frac{\min_{s\in R(x_k)}L_k(s)}
       {\min_{s\in\mathcal{S}}L_k(s)}\;\geq\;1 .
  \label{eq:router-latency-ratio}
\end{equation}
After routing, Eq.~\eqref{eq:selection}
chooses the valid candidate with the fewest cycles.  Neither branch sees
optimized-candidate outcomes at prediction time.

\subsection{Post-Training of Open Model}
\label{sec:model-adaptation-method}

In addition, we study two complementary forms of post-training for open models that serve as the base model of \system{}.
(1) Offline supervised fine-tuning (SFT) distills feasible actions from a stronger
teacher, whereas (2) reinforcement learning directly favors programs that obtain
higher measured HLS quality of results (QoR).

\subsubsection{Offline SFT from Validated Agent Actions}
We build supervision from expertise-guided teacher artifacts.  Each example
joins a model-call context $x$---agent role, HLS kernel, feedback, and indexed
expertise---with the response $y$ and the corresponding candidate's Vitis
evaluation.  An example enters the positive set $\mathcal{D}^{+}$ only if the
response contains code and the derived candidate completes C simulation and
HLS synthesis while meeting the target period and device constraints.

\begin{table*}[t]
  \centering
  \caption{Evaluation matrix.  HLS cycles are synthesis estimates; RTL cycles
  come from co-simulation.}
  \label{tab:experimental-setup}
  \scriptsize
  \setlength{\tabcolsep}{5pt}
  \renewcommand{\arraystretch}{1.12}
  \begin{tabularx}{\textwidth}{@{} l Y{0.80} Y{0.95} Y{1.25} @{}}
    \toprule
    \textbf{Study} & \textbf{Models / workload} &
    \textbf{Compared configurations} & \textbf{Evidence} \\
    \midrule
    Teacher sweep &
    Sonnet~4.6, Opus~4.8, GPT-5.5; 26 PolyBench kernels &
    Zero-shot; flash/multi-step $\times$ no, selected, or all expertise &
    C simulation, HLS synthesis, timing, and fit; no golden-output or RTL check \\
    \addlinespace
    Open-model sweep &
    Gemma-4-31B\cite{gemmateam2026gemma4technicalreport}, Qwen3.6-27B\cite{qwen3.6-27b}; 27 PolyBench kernels &
    flash/multi-step $\times$ five expertise policies &
    CPU-golden output, C simulation, HLS synthesis, timing, and fit \\
    \addlinespace
    Setup router &
    Sonnet~4.6; 19 group-held-out kernels &
    Predicted five versus exhaustive ten setups &
    Group-held-out ranking; valid-candidate HLS cycles \\
    \addlinespace
    Supervised fine-tuning &
    Qwen3.6-27B; five held-out kernels &
    Zero-shot; flash/multi-step $\times$ expertise off/on &
    Agentic: CPU golden through timing/fit; Zero-shot: print-only C simulation
    and synthesis; no RTL \\
    \addlinespace
    Policy optimization &
    Qwen2.5-Coder-7B; 14 training and 24 evaluation tasks &
    Base versus supervised versus reward-tuned &
    Training: C-simulation-gated QoR reward; evaluation: reported RTL \\
    \addlinespace
    Additional RTL &
    DeepSeek-v4 Flash \cite{xu2026deepseek}, Grok, HLSmith/ChatHLS &
    Expertise sensitivity and cross-system comparison &
    Supplied RTL cycles; no independent output replay \\
    \bottomrule
  \end{tabularx}
\end{table*}

The target model's chat template renders each conversation, and a completion
mask $M_{x,y,t}$ equals one only for the complete assistant target and excludes
the prompt and framework context.  
With frozen base-model parameters $\theta$ and trainable adapter parameters $\phi$,
the objective is as follows:
\begin{equation}
  \mathcal{L}_{\mathrm{SFT}}(\phi)=
  -\frac{\sum_{(x,y)\in\mathcal{D}^{+}}\sum_t
  M_{x,y,t}\log p_{\theta,\phi}(y_t\mid x,y_{<t})}
  {\sum_{(x,y)\in\mathcal{D}^{+}}\sum_t M_{x,y,t}}.
  \label{eq:sft-objective}
\end{equation}
Because each supervised completion token receives the same cross-entropy
weight and latency, neither weights nor selects examples, SFT imitates
validated, expertise-conditioned teacher actions without directly optimizing latency, neither invoking Vitis nor an external LLM.   
The teacher, target roles,
corpus split, and adapter settings appear in
Section~\ref{sec:experimental-setup}.

\subsubsection{Vitis-Grounded Group-Relative Policy Optimization}

To optimize QoR directly, we formulate Translator adaptation as a single-step contextual bandit rather than reinforcement learning over the multi-step agentic workflow.  The context is the plain source, header, and top-level function; an action is one complete generated HLS kernel.  
QoR from Vitis C simulation and synthesis are mapped to a scalar reward.  
The adapted Translator can subsequently be
used inside the full HLSmith workflow.  
We use group-relative policy optimization (GRPO)~\cite{shao2024deepseekmath}, summarized in
Algorithm~\ref{alg:vitis-grpo}.

$a_{k,i}$, $z_{k,i}$ is its terminal tool outcome and $Q_{k,i}$ contains latency $L_{k,i}$ and resource use $U_{k,i,d}$.  The reward
separates a correctness ladder from latency benefit and resource pressure:
\begin{equation}
\begin{aligned}
\widetilde r_{k,i}(h)={}&b+\alpha h\!\left(\frac{L_k^0}{L_{k,i}}\right)\\
&-\beta\sum_d\max\!\left(0,\frac{U_{k,i,d}}{B_d}-1\right).
\end{aligned}
\label{eq:grpo-reward}
\end{equation}
Here $d$ ranges over FPGA resource classes and $B_d$ is the corresponding hardware resource
scoped by the target FPGA.  
Equation~\eqref{eq:grpo-reward} gives latency credit only after the tool outcome reaches the valid-design tier; lower tiers receive terminal
scores.  Among valid candidates, $h$ rewards speedup relative to the baseline,
while the final term penalizes only resource-budget excess defined by the target FPGA.
Concrete reward constants,
rollout settings, adapter parameters, and training data appear in
Section~\ref{sec:experimental-setup}.

Table~\ref{tab:router-setup-space} gives the router's complete configuration
space.  Matched-positive, best-fit, and exhaustive are sub-policies of
\emph{Select expertise}; each exposes a kernel-dependent subset of the expertise library.



\section{Evaluation}
\subsection{Experimental Setup}
\label{sec:experimental-setup}

{
Table~\ref{tab:experimental-setup} provides an overview of \system{} experiment setup for evaluation.
The teacher-model sweep crosses flash and multi-step execution with no,
selected, or all architectural expertise.  Zero-shot generation is a separate arm.  Within a study, comparisons fix the model, input, initial design and report, prompt, and candidate hardware resource budget.  

Flash mode candidates independently
start from the same state; multi-step runs condition each action one by one on
the newly accepted tool result.  Exhaustive best-of-setup values are search ceilings.  
The learned router is evaluated
separately through Eq.~\eqref{eq:router-latency-ratio}.

\label{sec:evaluation}

\begin{figure*}[!t]
  \centering
  \includegraphics[width=0.98\textwidth]{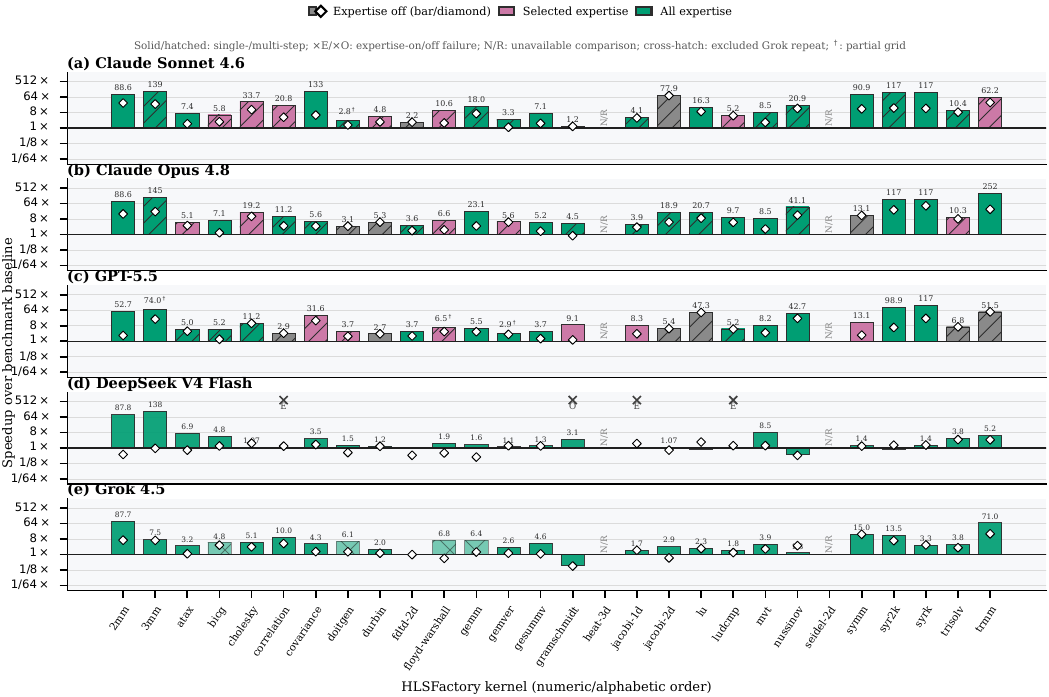}
  \caption{Per-kernel benchmark-baseline speedup.  Bars show $B_k/S_k$ and
  diamonds $B_k/O_k$.  Rows (a--c) use HLS synthesis; rows (d--e) use RTL
  co-simulation.  Crosses denote failures, daggers incomplete grids, N/R a
  missing baseline, and cross-hatch unmatched Grok reruns.}
  \label{fig:aligned-model-expertise}
\end{figure*}

\begin{table*}[!t]
  \centering
  \caption{Common-cohort summary of Fig.~\ref{fig:aligned-model-expertise}. Coverage is eligible kernels/26; teacher rows require all six setups, whereas RTL rows require an executed baseline and both flash arms. Five unmatched Grok reruns are excluded.}
  \label{tab:aligned-model-expertise}
  \scriptsize
  \setlength{\tabcolsep}{3.5pt}
  \renewcommand{\arraystretch}{1.00}
  \begin{tabular}{@{}lllcrrr@{}}
    \hline
    \textbf{Model} & \textbf{Evidence} & \textbf{Compared setups} & \textbf{Coverage} & \shortstack{\textbf{Benchmark-baseline}\\\textbf{speedup}} & \shortstack{\textbf{Expertise-off}\\\textbf{speedup}} & \shortstack{\textbf{Expertise}\\\textbf{wins/losses}} \\
    \hline
    Claude Sonnet 4.6 & HLS synthesis & Exhaustive setup selection & 25/26 & 17.46$\times$ & 2.88$\times$ & 22/3 \\
    Claude Opus 4.8 & HLS synthesis & Exhaustive setup selection & 26/26 & 14.51$\times$ & 2.63$\times$ & 23/3 \\
    GPT-5.5 & HLS synthesis & Exhaustive setup selection & 23/26 & 11.39$\times$ & 2.30$\times$ & 17/6 \\
    DeepSeek V4 Flash & RTL co-simulation & Flash mode & 22/26 & 2.58$\times$ & 2.55$\times$ & 17/5 \\
    Grok 4.5 & RTL co-simulation & Flash mode & 21/26 & 4.47$\times$ & 1.87$\times$ & 17/4 \\
    \hline
  \end{tabular}
\end{table*}

Motivated by the three major challenges identified in
Section~\ref{sec:introduction}---limited architectural reasoning in
general-purpose LLMs, insufficient procedural knowledge of HLS development,
and difficulty selecting kernel-appropriate optimization techniques---we
evaluate the effectiveness of expertise and workflow design, early
setup/configuration routing, and open models with and without post-training
using HLS synthesis, correctness checks, and,
ultimately, RTL co-simulation profiles.
The router validation split selects benchmarks by the architecture characteristics, performance and resources thresholds, and
the complexity of the benchmark.  In the validation split, selected kernels are \coden{durbin},
\coden{floyd\_warshall}, \coden{gemm}, and \coden{trmm}. 
For the GRPO experiment, we set that $b=0.2$, $\alpha=0.6$, $\beta=0.5$,
$h(z)=\min(2,z)$, $B_d$ to 50\% of U280 capacity (to help encourage feasible designs), and
$\epsilon_A=10^{-4}$ in Eq.~\eqref{eq:grpo-reward}.  Valid rewards are clipped
to $[-0.4,2]$; Abnormal cases such as empty or simulation-dependent output, tool failure or timeout,
and mismatch or near-empty circuits receive $-1.0$, $-0.8$, and $-0.6$,
respectively. The C simulation serves as the gate for rollouts; target timing and RTL correctness remain downstream evaluations.

HLSFactory, serving as our major benchmark, supplies the construction flow for 28 PolyBench HLS kernels
\cite{pouchet2012polybench,abikaram2024hlsfactory}. On top of that, output value checks are added to ensure correctness signals perform normally. This becomes "CPU-golden" as golden references. As a result of our added measures, table~\ref{tab:experimental-setup} establishes
independent numerical agreement.  Unless the table states otherwise, a
candidate is eligible only after C simulation, synthesis, target-timing, and
device-resource checks.  Throughout the experiment, we report valid-design counts and speedup ratios, with all
failures and exclusions
\cite{dolan2002performance}.   A common valid denominator is used to produce Geometric means of the overall speedup against golden references;
paired confidence intervals use bootstrap resampling
\cite{efron1994bootstrap}.  Vitis HLS 2023.2 targets
 AMD Alveo U280 FPGA (device: xcu280-fsvh2892-2L-e) at 300 MHz throughout the matched local studies.
}

\subsection{Architectural Expertise and Multi-Step Optimization}
\label{sec:eval-teacher}

For kernel $k$, let $B_k$, $S_k$, and $O_k$ be the benchmark-baseline, selected-design, and selected expertise-off cycles.
Table~\ref{tab:aligned-model-expertise} reports geometric means of $B_k/S_k$ and
$O_k/S_k$ on a common cohort; values above one favor $S_k$.  The agentic framework using teacher (Claude Sonnet 4.6) and other models (Claude Opus 4.8, Grok 4.5, and DeepSeek v4-flash) is tested against six agentic setups, with all three sub-categories of the select-expertise policy merged into a single class for notational simplicity, while RTL co-simulation rows use a single all-expertise mode setup.


Table~\ref{tab:aligned-model-expertise} and
Fig.~\ref{fig:aligned-model-expertise}(a--c) summarize three sweeps.
Sonnet~4.6 completes
\NewSkillCompletionCount{}/\NewSkillScheduledCount{} requested cells: all 26
zero-shot and \NewSkillAgenticCompletionCount{}/
\NewSkillAgenticScheduledCount{} agentic cells; two multi-step \coden{doitgen}
cells are absent.  All \NewSkillAgenticFeasibleCount{} completed agentic cells meet timing and resource limits, versus \NewSkillOneShotTimingCount{}/26
zero-shot designs.  On \NewSkillCommonKernelCount{} completed tasks, $B_k/S_k$ is
\NewSkillBestGmean$\times$ versus $B_k/O_k=$\NewSkillNoSkillGmean$\times$;
thus expertise adds \NewSkillIncrementalGmean$\times$ and wins
\NewSkillEnabledWinCount{}/\NewSkillCommonKernelCount{} kernels against expertise off.

In comparison, Opus 4.8 advances 15/26 cases in zero-shot and
25/26 agentic designs meet timing and resource limits.  
With expertise selection policy applied, an overall speedup of
$B_k/S_k=$\JulyThirtyOpusBestBaselineGain$\times$ and
$O_k/S_k=$\JulyThirtyOpusBestOverOffGain$\times$; are reached, and
 and beating expertise off cases in \JulyThirtyOpusGuidedSelections{}/26 benchmarks; expertise selection policy yields
\JulyThirtyOpusFlashSelectedOverOff$\times$/
\JulyThirtyOpusMultiSelectedOverOff$\times$ cycle reduction 
\JulyThirtyOpusFlashAllOverOff$\times$/
\JulyThirtyOpusMultiAllOverOff$\times$ for all expertise policy applied (flash/multi-step).  
Selected expertise
benefits more from rerouting, whereas all expertise helps more in one step.
\begin{figure*}[!t]
  \centering
  \includegraphics[width=0.98\textwidth]{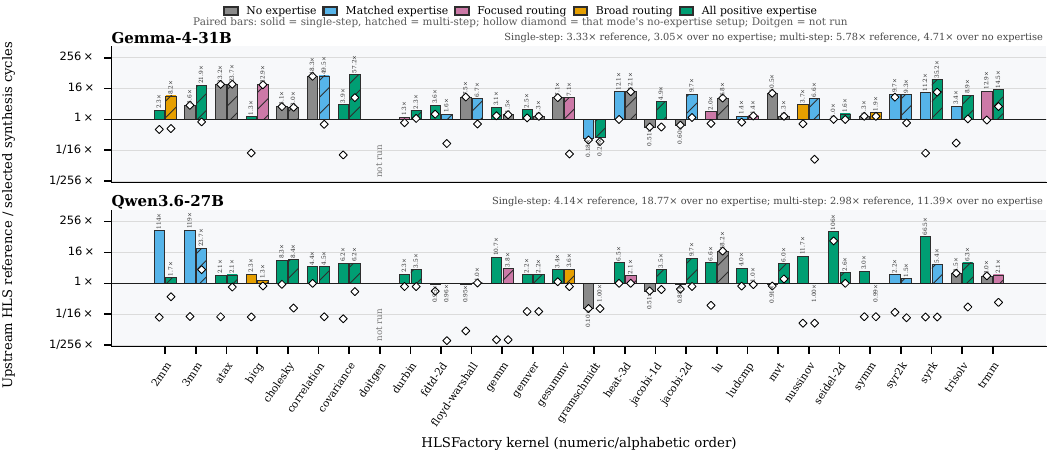}
  \caption{Open-weight expertise sweep.  Within each mode, bars show the best
  qualified speedup over the upstream reference across five policies; diamonds
  show expertise off.  Solid/hatched bars denote flash/multi-step execution.
  The 27 kernels pass CPU-golden, simulation, synthesis, timing, and resource
  checks.}
  \label{fig:skill-v3-open-models}
\end{figure*}
In addition, GPT-5.5 executes
\GPTFiftyFiveAgenticAttemptCount{}/\GPTFiftyFiveAgenticExpectedCount{} agentic
cells; \GPTFiftyFiveAgenticCSynthPassCount{} synthesize and
\GPTFiftyFiveAgenticFeasibleCount{} also meet timing and resource limits.  On
\GPTFiftyFiveCommonKernelCount{} complete grids, $B_k/S_k$ is
\GPTFiftyFiveBestBaselineGain$\times$, $O_k/S_k$ is
\GPTFiftyFiveBestOverOffGain$\times$, and expertise wins
\GPTFiftyFiveSkillsSelections{} kernels against expertise-off.  
\begin{table}[!t]
  \centering
  \caption{Five-of-ten setup routing on 19 group-held-out development kernels.
  Counts show inclusion of the exhaustive optimum or a workflow within 5\%.
  }
  \label{tab:setup-router}
  \scriptsize
  \setlength{\tabcolsep}{3pt}
  \renewcommand{\arraystretch}{1.1}
  \begin{tabularx}{\columnwidth}{@{} Y{1} c c c @{}}
    \toprule
    \textbf{Router} &
    \begin{tabular}[b]{@{}c@{}}\textbf{Exhaustive}\\\textbf{optimum}\end{tabular} &
    \begin{tabular}[b]{@{}c@{}}\textbf{Within 5\%}\\\textbf{of optimum}\end{tabular} &
    \begin{tabular}[b]{@{}c@{}}\textbf{Latency ratio}\\\textbf{geometric mean /}\\\textbf{95th percentile}\end{tabular} \\
    \midrule
    Shallow multilayer perceptron & 15/19 & 16/19 & 1.142$\times$ / 2.413$\times$ \\
    Pairwise ranking network & 11/19 & 13/19 & 1.180$\times$ / 2.206$\times$ \\
    Qwen embedding nearest neighbor & 16/19 & 16/19 & 1.081$\times$ / 1.642$\times$ \\
    Extremely randomized trees (pairwise) & \textbf{17/19} & 17/19 & 1.056$\times$ / 1.362$\times$ \\
    Prompted Qwen3-0.6B & 14/19 & \textbf{18/19} & 1.103$\times$ / 1.526$\times$ \\
    Fine-tuned Qwen3-0.6B & 8/19 & 9/19 & 1.887$\times$ / 12.703$\times$ \\
    \midrule
    HLSmith hybrid router (ours)$^{\dagger}$ & \textbf{17/19} & \textbf{18/19} & \textbf{1.026$\times$} / \textbf{1.087$\times$} \\
    \bottomrule
  \end{tabularx}
\end{table}
In short, the expertise-related framework configurations win most completed tasks for every API-called model, while
the Opus mode results show that its best exposure depends on workflow depth.
These results support explicit, kernel-dependent expertise rather than a
universally best setup.

In Figure \ref{fig:aligned-model-expertise}, with exhaustive router applied, rows (a--c) present the routed best per-benchmark configuration speedup over the benchmark reference HLS code
across all six major agentic setups.
Except for a few ties against the expertise-off cases in speedup ratios (e.g., 5 ties out of 26 benchmarks for \system{} with Claude Sonnet 4.6), our framework can adaptively select the optimal configurations across various tasks, with the obvious improvement in expertise adopted.

Even with flash mode implemented only, rows (d--e) 
DeepSeek and Grok reach
\JulyThirtyDeepSeekBaselineGain$\times$ and
\JulyThirtyGrokBaselineGain$\times$ over the benchmark
baseline on \JulyThirtyDeepSeekStrictCount{} and
\JulyThirtyGrokStrictCount{} kernels, respectively.  
On the
\JulyThirtyGrokDeepSeekCommonCount{} common enabled-arm kernels, Grok uses
\JulyThirtyGrokOverDeepSeekGain$\times$ fewer cycles.  
The within-model expertise
effect is distinct: it reduces cycles by
\JulyThirtyDeepSeekStrictExpertiseGain$\times$ for DeepSeek and
\JulyThirtyGrokStrictExpertiseGain$\times$ for Grok.
Grok is faster, whereas DeepSeek responds more to expertise. 

Requiring only both generated arms to pass leaves
\JulyThirtyDeepSeekPairedCount{} DeepSeek pairs and
\JulyThirtyGrokPrimaryCount{} Grok pairs; expertise reduces cycles by
\JulyThirtyDeepSeekGain$\times$ and \JulyThirtyGrokPrimaryGain$\times$,
with wins/losses of
\JulyThirtyDeepSeekWins{}/\JulyThirtyDeepSeekLosses{} DeepSeek pairs and
\JulyThirtyGrokPrimaryWins{}/\JulyThirtyGrokPrimaryLosses{} Grok pairs.
For five Grok kernels, only the expertise-enabled arm receives an additional
attempt; cross-hatching marks these excluded reruns.  The CSVs omit generated
RTL and output traces, preventing independent value replay.
\begin{figure}[!t]
  \centering
  \includegraphics[width=\columnwidth]{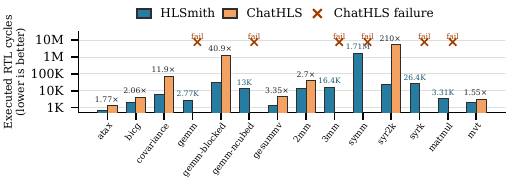}
  \caption{RTL outcomes for 14 attempts.  Labels show
  ChatHLS/HLSmith speedup on paired passes.  Crosses mark
  \JulyThirtyChatHLSFailureCount{} ChatHLS failures; blue labels give HLSmith
  cycles.  Outputs are not replayed.}
  \label{fig:chathls-cosim}
\end{figure}
Figure~\ref{fig:skill-v3-open-models} 
shows that all \SkillVThreeOpenCellCount{} outputs also pass C simulation, synthesis, timing, and resource gates;
\SkillVThreeOpenBaselineFallbackCount{} retains the translated baseline after
unsuccessful edits.  

For Gemma, it reaches
\SkillVThreeOpenGemmaFlashReferenceReduction$\times$/
\SkillVThreeOpenGemmaMultistepReferenceReduction$\times$ relative to the
upstream HLS reference and
\SkillVThreeOpenGemmaFlashOracleOverNoExpertise$\times$/
\SkillVThreeOpenGemmaMultistepOracleOverNoExpertise$\times$ relative to the
no-expertise arm (flash/multi-step).  Qwen3.6-27B\cite{qwen3.6-27b} reaches
\SkillVThreeOpenQwenFlashReferenceReduction$\times$/
\SkillVThreeOpenQwenMultistepReferenceReduction$\times$ and
\SkillVThreeOpenQwenFlashOracleOverNoExpertise$\times$/
\SkillVThreeOpenQwenMultistepOracleOverNoExpertise$\times$.  Expertise-enabled
policies supply the minimum on
\SkillVThreeOpenGemmaFlashExpertiseEnabledSelections{}/\SkillVThreeOpenKernelCount{}
and \SkillVThreeOpenGemmaMultistepExpertiseEnabledSelections{}/
\SkillVThreeOpenKernelCount{} Gemma cases, and
\SkillVThreeOpenQwenFlashExpertiseEnabledSelections{}/\SkillVThreeOpenKernelCount{}
and \SkillVThreeOpenQwenMultistepExpertiseEnabledSelections{}/
\SkillVThreeOpenKernelCount{} Qwen cases.
These are post-hoc setup ceilings; every reference speedup divides the supplied
upstream HLS implementation's cycles by generated-design cycles, not by an
expert frontier.


We use ChatHLS as the quantitative comparison because it most closely matches HLSmith’s end-to-end scope, including HLS-C generation, tool-grounded repair, QoR-driven optimization, C/RTL verification, and HLS-specific model adaptation (\Cref{tab:system-scope}). The comparison between ChatHLS and our \system{} covers
\JulyThirtyChatHLSDenominator{} kernels attempted by both systems.  Overall, our HLSmith
reports \JulyThirtyHLSmithPass{} RTL co-simulation passes and ChatHLS
\JulyThirtyChatHLSPass{}.  On \JulyThirtyChatHLSPaired{} common passes, every
speedup ratio favors our HLSmith and
$\operatorname{GM}(L_{\mathrm{ChatHLS}}/L_{\mathrm{HLSmith}})=$
\JulyThirtyChatHLSGain$\times$.  
The large benchmark, \coden{syr2k}, whose speedup ratio is influential:
excluding it gives \JulyThirtyChatHLSWithoutSyrTwoKGain$\times$ on the other
The
\JulyThirtyHLSmithOnlyPass{} HLSmith-only passes improve recorded solve rate.


\subsection{Pruning Workflow Configurations with the Early Router}
\label{sec:eval-router}

Kernel-dependent winners motivate the outer router with early prediction functionality
(Section~\ref{sec:setup-router}), which can retain five of the ten setups per benchmark in
Table~\ref{tab:router-setup-space} and \RouterDevelopmentKernelCount{}
group-held-out development kernels.  During the training and evaluation, tt sees only source structure and the
initial HLS report; Sonnet~4.6 generates candidates.
Equation~\eqref{eq:router-latency-ratio} measures quality relative to exhaustive
selection, independently of the per-turn Action Router.
Table~\ref{tab:setup-router} shows that four structured
choices plus one semantic choice preserve ExtraTrees'
\RouterExtraTreesExactCount{}/19 exact-optimum coverage, increase
within-5\% coverage from \RouterExtraTreesWithinFiveCount{}/19 to
\RouterFusionWithinFiveCount{}/19, and reduce the 95th-percentile latency ratio
from \RouterExtraTreesTailRegret$\times$ to
\RouterFusionTailRegret$\times$ while pruning \RouterFusionPrunedPercent\% of
setups.  Pairwise fine-tuning improves token classification but worsens
synthesized latency.  

\subsection{Transferring Optimization Behavior to Open Models}
\label{sec:eval-sft}


\noindent\textbf{Local open-model configurations.}
Figure~\ref{fig:open-model-sft} compares Gemma-4-31B base with two rank-16
Qwen3.6-27B adapters on five kernels.  All \ModelMatrixAgenticValidCount{}
agentic outputs pass CPU-golden, C simulation, synthesis, timing, and resource
gates. With all expertise enabled, Gemma achieves
$0.50\times/1.02\times$ speedup over the recorded gold reference in
Flash/multi-step mode, respectively; the corresponding results are
$0.65\times/1.38\times$ for active-role fine-tuning and
$1.13\times/2.24\times$ for Orchestrator fine-tuning; expertise is therefore not
uniformly beneficial. The modes use two/five model calls.  Direct zero-shot C simulation and synthesis pass
\GemmaDirectCSynthCount{}/\GemmaDirectKernelCount{} Gemma-base,
\QwenBaseDirectCSynthCount{}/\QwenBaseDirectKernelCount{} Qwen-base, and 5/5
adapter kernels. 

\begin{figure}[!t]
  \centering
  \includegraphics[width=\columnwidth]{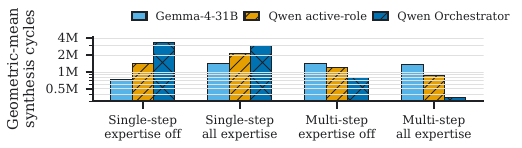}
  \caption{Five-kernel open-model pilot.  Bars show geometric-mean HLS
  synthesis cycles (lower is better) across workflow--expertise settings;
  color and hatch identify the model.  All outputs pass CPU, timing, and
  resource checks.  }
  \label{fig:open-model-sft}
\end{figure}

\noindent\textbf{Matched Translator adaptation diagnostic.}
A four-kernel A/B changes only the Qwen3.6-27B \cite{qwen3.6-27b} Translator; all other roles,
flash execution, expertise-off policy, temperature, seed, candidate
count, and Vitis setup remain fixed.  Fine-tuning reduces geometric-mean
synthesis and RTL-simulation cycles by \TranslatorSFTABHLSGain$\times$ and
\TranslatorSFTABRTLGain$\times$, improving two kernels and tying two; timing
passes change from
\TranslatorSFTABBaseTimingCount{}/\TranslatorSFTABKernelCount{} to
\TranslatorSFTABTunedTimingCount{}/\TranslatorSFTABKernelCount{}.
The four-pair RTL-ratio bootstrap interval,
[\TranslatorSFTABRTLCILow$\times$, \TranslatorSFTABRTLCIHigh$\times$], includes
one.  Table~\ref{tab:qwen-translator-sft-ab} gives per-kernel results. In short, with proper GRPO configuration, small models with as few as 7B parameters can achieve decent optimization within our agentic framework.

\begin{table}[!t]
  \centering
  \caption{Matched Qwen3.6-27B translator fine-tuning diagnostic.  Gains divide base by fine-tuned worst-case cycles.
  $\dagger$ marks a base target-clock miss.  
  }
  \label{tab:qwen-translator-sft-ab}
  \scriptsize
  \setlength{\tabcolsep}{3pt}
  \renewcommand{\arraystretch}{1.1}
  \begin{tabularx}{\columnwidth}{@{} Y{1} c c c @{}}
    \toprule
    \textbf{Kernel} &
    \begin{tabular}[b]{@{}c@{}}\textbf{HLS-estimate}\\\textbf{gain}\end{tabular} &
    \begin{tabular}[b]{@{}c@{}}\textbf{Executed RTL}\\\textbf{gain}\end{tabular} &
    \begin{tabular}[b]{@{}c@{}}\textbf{Fine-tuned}\\\textbf{selection}\end{tabular} \\
    \midrule
    \coden{durbin} & 1.00$\times$ & 1.00$\times$ & baseline retained \\
    \coden{floyd-warshall} & 1.00$\times$ & 1.00$\times$ & baseline retained \\
    \coden{gemm}$^{\dagger}$ & 3.61$\times$ & 2.91$\times$ & generated edit \\
    \coden{trmm} & 2.34$\times$ & 2.26$\times$ & generated edit \\
    \midrule
    Geometric mean & 1.70$\times$ & 1.60$\times$ & --- \\
    \bottomrule
  \end{tabularx}
\end{table}

\noindent\textbf{Common-task 7B comparison.}
Algorithm~\ref{alg:vitis-grpo} adapts Qwen2.5-Coder-7B using the earlier
linear, 2$\times$-capped reward (Section~\ref{sec:model-adaptation-method}).
SFT and GRPO use different training corpora but share
\GRPOMatchedKernelCount{} evaluation tasks.  RTL-output passes rise from
\GRPOBaseRTLPassCount{} for the base model to \GRPORTLPassCount{} after
group-relative policy optimization; SFT reaches \GRPOSFTRTLPassCount{}.  The
reported \GRPOMedianDescriptiveRatio$\times$ median ratio compares different
C-simulation-passing subsets and is neither paired nor a geometric mean.

\noindent\textbf{Post-training behavior.}
Across 26 base/reward-tuned tasks, \GRPOBrokenToCorrectCount{} change from
broken to RTL-correct, \GRPOBothCorrectFasterCount{} pass in both cases and
improve, and \GRPOBaseCorrectRegressionCount{} regress.  Inspection finds
loop fission, outward pipelining, reduction unrolling, and feeding-array
partitioning, but no dataflow, double buffering, tiling, or AXI-burst
restructuring.  On \coden{gesummv}, cycles fall from the
\GRPOGesummvReferenceCycles{}-cycle reference estimate to
\GRPOGesummvCycles{} (\GRPOGesummvGain$\times$).
\section{Conclusion}
\label{sec:conclusion}
HLSmith frames C/C++-to-HLS optimization as hardware-architecture design rather than code generation alone. It distills multi-stage optimization practices from expert HLS implementations into reusable architectural expertise and applies them through compiler-guided single- and multi-step workflows. Across four commercial teacher models, expertise-enabled configurations produce the best synthesized design for most kernels and achieve up to 252$\times$ speedup over the supplied HLS baselines. In executed RTL comparisons, HLSmith completes all 14 kernels, versus 57\% for ChatHLS, and delivers a 6.9$\times$ geometric-mean speedup over the eight common passes (4.2$\times$ on the seven-kernel sensitivity set). The hybrid router evaluates five of ten configurations while retaining a result within 5\% of exhaustive selection on 18 of 19 development kernels. Teacher-derived fine-tuning improves geometric-mean synthesis and RTL performance by 1.70$\times$ and 1.60$\times$, respectively, with a peak per-kernel synthesis gain of 3.61$\times$. These results demonstrate that explicit architectural expertise, compiler feedback, and targeted model adaptation enable agents to translate plain C/C++ into efficient HLS accelerators.

\endgroup

\typeout{HPCA-CONTENT-LAST-PAGE=\thepage}
\label{hpca:lastcontentpage}
\bibliographystyle{IEEEtranS}
\bibliography{refs}

\clearpage
\appendices
{\color{llmgray}\section{AI Use}
\label{app:ai-use}

OpenAI Codex materially assisted this paper by reviewing the implementation
and experimental results, identifying confounds, drafting and revising the
dark-gray text, and helping generate tables and figures from recorded
measurements.  It also assisted with consistency, citation, build, anonymity,
and evidence-boundary checks.  Codex did not create experimental measurements;
results with incomplete controls or correctness evidence remain labeled as
preliminary.

Hosted models used to generate HLS programs or training examples are identified
in the methodology and are treated as experimental subjects.  The human authors
define the research questions, design the system and evaluation, verify the
measurements and technical claims, approve the final text and figures, and
retain responsibility for the submission.  Dark gray marks Codex-authored text
during collaborative review and is removed after human approval.
}

\end{document}